\documentclass[11pt]{article}
\usepackage{moriond,epsfig}

\usepackage{amsmath}
\usepackage{graphicx}
\usepackage{graphics}
\usepackage{longtable}
\usepackage[dvips]{color}
\usepackage{color}
\usepackage{amssymb,amsfonts}
\usepackage{latexsym}

\newcommand{\hegra}{HEGRA}
\newcommand{\hess}{H.E.S.S.}
\newcommand{\cangaroo}{CANGAROO}
\newcommand{\magic}{MAGIC}
\newcommand{\veritas}{VERITAS}
\newcommand{\milagro}{Milagro}

\newcommand{\egret}{{\it EGRET}}
\newcommand{\fermi}{{\it Fermi}}

\newcommand{\un}[1]{~\hspace{-1pt}\ensuremath{\mathrm{#1}}}

\newcommand{\xray}{X-ray}
\newcommand{\xrays}{X-rays}

\newcommand{\pidec}{$\pi^{0}$-decay}

\newcommand{\rxj}{RX~J1713.7-3946}
\newcommand{\velajr}{RX~J0852-4622}
\newcommand{\rcw}{RCW~86}

\def\Journal#1#2#3#4{{#1} {\bf #2}, #3 (#4)}

\newcommand\araa{{ARA\&A}}%
\newcommand\apj{{ApJ}}%
\newcommand\apjs{{ApJS}}%
\newcommand\aap{{A\&A}}%
\newcommand\mnras{{MNRAS}}%
\newcommand\nat{{Nature}}%
\def\be{\begin{equation}}
\def\ee{\end{equation}}
\def\bea{\begin{eqnarray}}
\def\eea{\end{eqnarray}}

\def\edot{$\dot{{\rm E}}$}

\def\d{$^\circ$}
\def\m{$^\prime$}
\def\s{$^{\prime\prime}$}

\def\cm3{cm$^{-3}$~}

\def\eg{{\it e.g.~}}
\def\etal{et~al.~}
\def\ie{{\em i.e.~}}

\begin{document}
\vspace*{4cm}
\title{Latest results on Galactic sources as seen in VHE gamma-rays}

\author{M. Renaud}

\address{Laboratoire APC, CNRS-UMR 7164, Universit\'e Paris 7, \\
         10, rue Alice Domon et L\'eonie Duquet, 75025 Paris Cedex 13, France}

\maketitle
\abstracts{As of early 2009, latest results on Galactic sources (mainly shell-type and 
plerionic supernova remnants), as observed in the very-high-energy $\gamma$-ray domain, 
are reviewed. A particular attention is given to those obtained with the \hess~experiment 
during its Galactic Plane Survey which now covers the inner part of the Milky Way. From 
the well identified $\gamma$-ray sources to those without any obvious counterpart and 
the putative Galactic diffuse emission, this observational window fully deserves to be 
celebrated during this International Year of Astronomy, as a new mean to image the Galaxy 
and reveal sites of particle acceleration, potentially at the origin of Galactic cosmic rays.}

\section{Introduction}
\label{s:intro}

\begin{figure}[!htb]

\centering
\psfig{figure=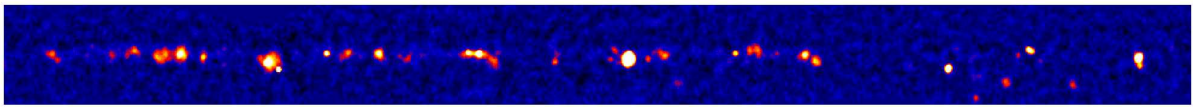,width=\textwidth}

\caption{\hess~significance image of the inner part of the Galaxy ($\ell$ $\in$ 
$[$-60\d,40\d$]$, $|$b$|$ $<$ 3\d), as of 2008 (from Chaves et al., \hess~collaboration, 
2008). The color transition from blue to red is set to 5 $\sigma$. \label{f:ima1}}

\end{figure}

\begin{figure}[!htb]

\centering
\psfig{figure=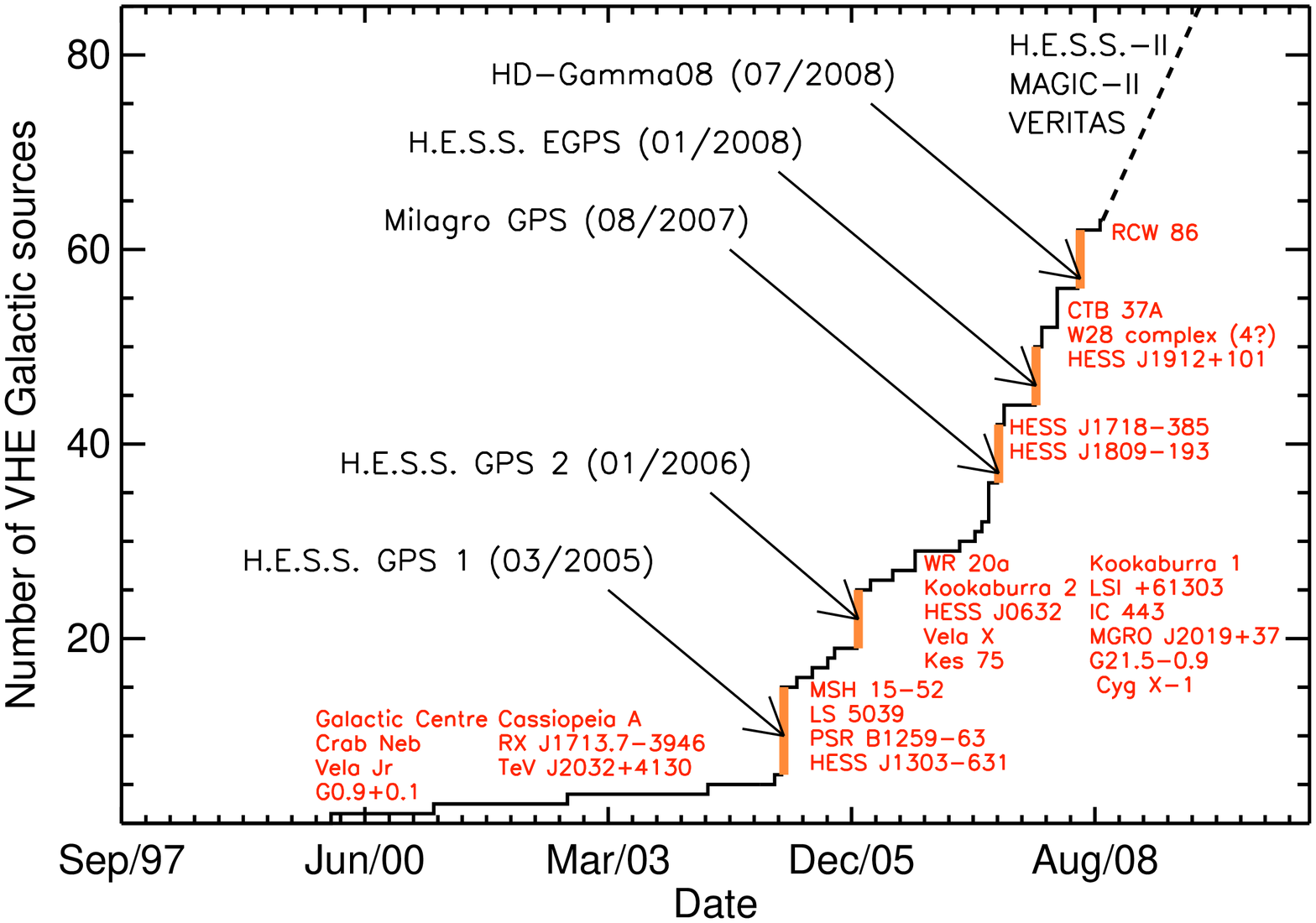,width=11.4cm}

\caption{A history of VHE Galactic astronomy. The number of VHE Galactic sources 
is shown against the date, which corresponds to either the date of the publication 
or the date of the conference where the discovery has been officially announced. 
The number of Galactic sources has tremendously increased by steps over the last 
five years, particularly through the Galactic Plane Surveys (GPS, EGPS standing for the 
extension of the GPS) conducted with \hess. HD-Gamma08 corresponds to the 4$^{th}$ 
International Meeting on High Energy Gamma-Ray Astronomy which was held in Heidelberg, 
july 2008. The sources marked in red have been revealed in between these steps, mainly 
thanks to dedicated observations. The extrapolation in time depicted by the dashed line 
serves as an estimate of the number of new sources which might be revealed in the incoming 
era of \hess~II (Southern Hemisphere), \magic~II and \veritas~(Northern Hemisphere).
\label{f:ima2}}

\end{figure}

Current generation of Imaging Atmospheric Cherenkov Telescopes (hereafter, IACTs)
have recently revealed a new population of Galactic sources emitting in the 
very-high energy (VHE; E $>$ 100\un{GeV}) gamma-ray domain \cite{c:gps06}. In
particular, the \hess~experiment, through a Galactic Plane Survey performed over 
the last five years and covering the inner Galaxy ($\ell$ $\in$ [-90\d,60\d], $|$b$|$ 
$<$ 3\d, see Figure \ref{f:ima1}), has accomplished a major breakthrough by revealing 
most of these new VHE $\gamma$-ray sources, as shown in Figure \ref{f:ima2}. A variety of 
source classes, identified (\ie coincident) with sources known at traditional wavelengths, 
was found, among them several shell-type supernova remnants (hereafter, SNRs), isolated 
or interacting with the surrounding medium, many young and middle-aged offset pulsar 
wind nebulae (hereafter, PWNe), some young massive star clusters, and a bunch of 
$\gamma$-ray binaries. In regards with SNRs and PWNe, one of the main pending question 
concerns the nature of the observed VHE $\gamma$-ray emission, which relates to the parent 
population of accelerated particles, or, in other words, the difficulty in disentangling 
the hadronic and leptonic contributions to the observed emission. This is in turn intimately 
linked to the more general question of the origin of Galactic cosmic rays (hereafter, CRs). 
As we shall see in the following, these questions can be efficiently addressed through a 
detailed investigation of the broadband spectrum of these sources, from radio to VHE 
$\gamma$-ray domains, coupled with the recent theoretical developments of acceleration 
mechanisms.

Besides these sources whose nature is firmly established thanks to the existing multi-wavelength 
observations, many others fall into the category of the so-called dark sources (\ie with no 
clear counterpart at other wavelengths) \cite{c:dark08}. This can be first explained by the fact 
that the majority of the VHE $\gamma$-ray sources are extended, on scales of the order of tens of arcminutes, 
with no clear sub-structure. Although current IACTs have reached unprecedented sensitivities 
and angular resolutions, the morphology of most of the faint sources can not be characterised 
precisely. Moreover, instruments in other domains (radio, infrared, \xrays) usually feature 
angular resolutions at the arcsecond / subarcminute scales, often coupled with relatively 
small field of views, which (1) does not permit one to perform deep surveys of the whole 
Galactic Plane, and (2) makes it difficult to reveal large-scale structures coincident with 
these newly discovered sources.

In this contribution, latest results on VHE Galactic sources are reviewed, with a particular
attention to those obtained with \hess. The well-identified cases, such as shell-type SNRs 
and PWNe, are discussed in sections \ref{s:snrs} and \ref{s:plerion} respectively, together 
with the implications and new questions related to the acceleration mechanisms and the nature 
of the VHE $\gamma$-ray emission. Some interesting cases of dark sources are exposed in section \ref{s:dark}, 
and the origin of the putative VHE Galactic diffuse emission is discussed in section \ref{s:diffuse}. 
The conclusion focuses on the interest of population studies, through the log N -- log S distribution 
of all the VHE Galactic sources known so far, and the perspectives with the next generation 
of IACTs (CTA \cite{c:cta}, AGIS \cite{c:agis}).

\section{Shell-type supernova remnants}
\label{s:snrs}

Since the first speculation of Baade \& Zwicky in 1934, the question of SNRs as the main sources
of Galactic CRs up to the knee ($\sim$ 3 PeV) and beyond is not yet settled, in spite 
of several decades of important observational \cite{c:reynolds08} and theoretical investigations 
\cite{c:malkov01,c:bell04,c:zp08}. The broadband spectrum of these sources, from radio to VHE $\gamma$-ray
domains, is the signature of particles accelerated at the shock fronts and radiating photons through 
several channels (synchrotron SC, non-thermal bremsstralhung, inverse-Compton IC and \pidec). Therefore, 
it represents by far our best access to the acceleration processes in SNRs \cite{c:ellison07}.
Up to now, five previously known shell-type SNRs, namely Cas~A~\cite{c:casA01,c:casA07}, 
\rxj~\cite{c:j1713_1,c:j1713_2,c:j1713_3}, \velajr~\cite{c:katagiri05,c:velajr1,c:enomoto06,c:velajr2}, 
\rcw~\cite{c:rcw86}, and more recently SN~1006 \cite{c:sn1006_2}, have been discovered in VHE $\gamma$-rays, 
the last four exhibiting a shell-type morphology matching that observed in \xrays. As discussed before, 
the main question concerns the nature of the VHE $\gamma$-ray emission from these shell-type SNRs (leptonic through
IC emission or hadronic through \pidec~decay from p-p interactions) and gives rise to an intense debate 
\cite{c:uchiyama07,c:vkb08,c:bv08,c:ellison08,c:butt08,c:plaga08}. On one hand, the correlation between 
\xray~and VHE $\gamma$-ray emission would support the leptonic scenario but implies, in a simple one-zone 
model, a spatially-averaged magnetic field of the order of a few tens of $\mu$G (except Cas~A). This 
value seems uncomfortably low in comparison with the theoretical prediction of magnetic field 
amplication associated with the efficient production of CRs at forward shocks, by the (non-linear) 
diffusive shock acceleration (DSA) mechanism \cite{c:malkov01,c:amato08}. Magnetic fields of $\gtrsim$ 
100 $\mu$G have been derived from the measured thickness of the \xray~filaments in several 
young SNRs \cite{c:vbk05,c:parizot06}, and, more recently, from the fast variability of small-scale \xray~dots 
and clumps \cite{c:uchiyama07}, in case these localized structures effectively reflect the SC losses of high-energy 
electrons in strong (amplified) magnetic fields \footnote{However, it was proposed that the thickness 
of the \xray~filaments would trace the magnetic field damping downstream of the forward shock and, therefore, 
would not be a measure of the magnetic field strength \cite{c:pohl05}. As for dots and clumps, which would 
reflect the inherent turbulence of the magnetic field, even in the case of a steady particle distribution 
 \cite{c:bykov08}.}. On the other hand, for the four resolved shell-type SNRs, the lack of clear 
correlation between the tracers of the ISM and the VHE $\gamma$-ray emission, together with the tight 
constraints on the local density derived from the absence (or the faint level) of thermal \xray~emission 
\cite{c:slane01,c:acero07}, does not permit one to draw firm conclusions in favor of an hadronic origin in 
these shell-type SNRs \footnote{Note that, in the framework of the non-linear DSA, the post-shock gas temperature, 
which is expected to lie in the \xray~domain for young SNRs, and consequently the thermal \xray~emission, can 
be reduced (shifted towards lower temperatures) due to strong particle acceleration \cite{c:drury09}. 
However, there are still discrepancies on the level at which the thermal \xray~emission could be suppressed 
\cite{c:patnaude09}.}.

\begin{table}[!htb]

\caption{Observational constraints on VHE shell-type SNRs. Distances d$_{\textrm{\tiny SNR}}$
assumed here are 1, 1, 1, 2.2, 3.4, 4.8 and 2.3\un{kpc}, for RX~J1713, Vela~Jr, \rcw, SN~1006, 
Cas~A, Kepler and Tycho, respectively. The first column gives the magnetic field 
values assuming a one-zone leptonic model, with standard seed radiation fields (CMB and Galactic 
infra-red and star-light emissions). The second column shows the magnetic field values derived from 
the thickness of the \xray~filaments. Third and fourth colums give the widths of VHE shells 
and of \xray~filaments, respectively. $\eta_{\textrm{\tiny CR}}$ in the sixth column represents the 
fraction of the energy of the explosion, E$_{\textrm{\tiny 51}}$ (in units of 10$^{51}$ erg), injected 
into CR protons, at the distance d$_{\textrm{\tiny SNR}}$ and for a gas density n$_{\alpha}$ = n/$\alpha$ 
cm$^{-3}$. Such density can then be compared to those given in the last column, mainly derived from 
the level of thermal \xray~emission. \label{t:snrs}}

\footnotesize

\vspace{0.4cm}
\begin{center}
\begin{tabular}{|c|c|c|c|c|c|c|}
\hline

 & B$_{\textrm{\tiny X-VHE}}$ &        B$_{\textrm{\tiny filaments}}$           &    Width$_{\textrm{\tiny VHE}}$ &  Width$_{\textrm{\tiny filaments}}$      &        $\eta_{\textrm{\tiny CR}}$      & n$_{\textrm{\tiny obs}}$   \\
 &  ($\mu$G)   & [$\times$d$_{\textrm{\tiny d}_{{\textrm{\tiny SNR}}}}^{-2/3}$] ($\mu$G) & [$\times$d$_{\textrm{\tiny d}_{{\textrm{\tiny SNR}}}}$] (pc) & [$\times$d$_{\textrm{\tiny d}_{{\textrm{\tiny SNR}}}}$] (pc) & [$\times$E$_{\textrm{\tiny 51}}$d$_{\textrm{\tiny d}_{{\textrm{\tiny SNR}}}}^{2}$] & (cm$^{-3}$) \\ 

\hline
\hline
RX~J1713 & $\sim$ 10  & 58--271  &      $\sim$ 4.5     & 0.1--0.2   & 0.8--2.6/n$_{0.1}$    & $<$ 0.02 d$_{\textrm{\tiny d}_{{\textrm{\tiny SNR}}}}^{-1/2}$   \\
Vela~Jr  & $\sim$ 6   & 200--240 &      2.2--3.9       & 0.18--0.44 & $\sim$ 2.5/n$_{0.1}$  & $<$ 0.03 d$_{\textrm{\tiny d}_{{\textrm{\tiny SNR}}}}^{-1/2}$   \\
\rcw     & $\sim$ 30  & 50--115  &      2.2--4.7       & 0.29--0.5  & 0.05--0.3/n$_{0.5}$   & (0.3--0.7)$_{\textrm{\tiny N}}$, $\sim$10$_{\textrm{\tiny S}}$          \\
SN~1006  & $\sim$ 30  & 57--143  &      2.4--3.5       & 0.13--0.2  & $\sim$ 0.2/n$_{0.05}$ & 0.05$_{\textrm{\tiny SE}}$, $\sim$ 0.2$_{\textrm{\tiny NW}}$             \\
Cas~A    & $\sim$ 100 & 485--550 & \textrm{unresolved} & 0.03--0.05 & $\sim$ 0.01/n$_{11}$  & $\sim$ 11$_{\textrm{\tiny shocked shell}}$ \\
Kepler   & $>$ 70     & 172--258 &         --          & 0.07--0.11 & $<$ 0.05/n$_{0.7}$    & $<$ 0.15$_{\textrm{\tiny SE}}$             \\
Tycho    & $>$ 70     & 240--360 &         --          & 0.04--0.05 & $<$ 0.02/n$_{0.4}$    & $<$ 0.3$_{\textrm{\tiny SC rim}}$  \\

\hline
\end{tabular}
\end{center}
\end{table}

Table \ref{t:snrs} summarizes the relevant parameters of the five shell-type SNRs detected in VHE 
$\gamma$-rays and of the two historical SNRs, Kepler and Tycho, for which upper limits have been 
obtained so far \cite{c:tycho01,c:kepler08}. It appears clearly that the magnetic field values estimated 
from a one-zone leptonic model (f$_{\textrm{\tiny X}}$/f$_{\textrm{\tiny VHE}}$ $\propto$ B$^{2}$), 
are significantly below those derived from the thickness of the \xray~filaments. The situation 
looks even worse in the case of Kepler and Tycho SNRs, for which only rather strongly amplified 
magnetic fields seem to be compatible with the non-detection of VHE $\gamma$-rays, within these simple 
assumptions \cite{c:vkb08}. On the other hand, the typical width of the \xray~filaments, over which 
amplified magnetic fields of about hundreds of $\mu$G may exist, are an order of magnitude below the widths
of the resolved VHE $\gamma$-ray shells. Hence, in case the magnetic field has been damped quickly behind 
the forward shock, the observed emission could still be explained by IC emission, with a fairly weak 
{\it spatially-averaged} value. Moreover, a detailed modeling of the interstellar radiation field 
for the calculation of the IC spectrum may help to improve the fit to the VHE $\gamma$-ray spectra \cite{c:porter06}. 
In regards with the hadronic scenario, the energy injected into CR protons, required to explain the 
VHE $\gamma$-ray flux, is quite demanding, especially if the constraints on the gas density from the 
level of thermal \xray~emission are taken at face value. Apart from these energetical considerations, 
\pidec~from p-p interactions seems to better explain the highest energy ($>$ 10\un{TeV}) data points 
measured in \rxj, where the Klein-Nishina limit of IC scattering takes place \cite{c:morlino09}. 
Therefore, both scenarios, in their simpliest form, suffer from severe limitations. The question of 
these shell-type SNRs as efficient Galactic CR accelerators can only be efficiently addressed 
through spectro-imaging analysis in \xrays~and VHE $\gamma$-rays, two domains whose current 
instrumental characteristics are quite different \footnote{While soft \xray~($<$ 10\un{keV}) telescopes
feature angular resolutions at the arcsecond scales, both the imaging instruments above 10\un{keV} 
and current IACTs reach angular resolutions of the order 5-10\m~at best.}, together with theoretical 
developments of the DSA mechanism.

\section{Plerionic supernova remnants}
\label{s:plerion}

Besides shell-type SNRs, a significant fraction of VHE $\gamma$-ray sources is (or at least seems to 
be) associated with energetic pulsars \cite{c:gallant08,c:pwne08}. These sources can generate 
bubbles of relativistic particles and magnetic field when their ultra-relativistic wind interacts 
with the surrounding medium (SNR or interstellar medium) \cite{c:gs06}. Their confinement leads to 
the formation of strong shocks, which can accelerate particles up to hundreds of TeV and beyond, thus
generating luminous nebulae seen across the entire electromagnetic spectrum in the SC emission from 
radio to hard \xrays, and through IC process and potentially \pidec~from p-p interactions \cite{c:horns06}, 
in the VHE $\gamma$-ray domain. On one hand, recent advances in the study of PWNe have been made from 
mainly radio and \xray~observations of the complex morphology of the inner PWN structure at the arcsecond 
scales \cite{c:gs06}. On the other hand, in the VHE domain, \hess~has proven to be capable to measure, in 
at least one case \cite{c:j1825}, spatially resolved spectra at the tenths of degree scales. These
complementary VHE observations then permit one to probe the electron spectra in these sources and 
to investigate the associated magnetic field distribution \cite{c:dejager08}.

Two classes of VHE $\gamma$-ray PWNe have recently emerged, based on observational grounds: young systems such 
as the Crab nebula \cite{c:crab}, G0.9+0.1 \cite{c:g09}, MSH 15-52 \cite{c:msh1552} and the newly discovered
VHE $\gamma$-ray sources associated with the Crab-like pulsars of G21.5-0.9 \cite{c:youngpwne} and Kes~75 \cite{c:kes75}, 
and evolved (extended and resolved) systems (\ie with characteristic ages $\tau_{c}$ $\gtrsim$ 10$^{4}$ yr), 
as exemplified by Vela~X \cite{c:velax}, HESS~J1825-137 \cite{c:j1825}, HESS~J1718-385 and 
HESS~J1809-193 \cite{c:twopwne}. In the former case, the VHE $\gamma$-ray emission, when resolved, matches quite 
well the morphology seen in \xrays, while in the latter case, these VHE $\gamma$-ray PWNe were found to be significantly 
offset from the pulsar position, with large size ratios between the \xray~and VHE $\gamma$-ray emission regions. 
The evolution of the SNR blastwave into an inhomogeneous ISM \cite{c:blondin01} and/or the high velocity of
the pulsar \cite{c:vds04}, together with a low magnetic field value ($\sim$ 5 $\mu$G \cite{c:dejager08}), 
may explain these large offset filled-center VHE $\gamma$-ray sources as being the relic nebulae from the past 
history of the pulsar wind inside its host SNR. Since VHE-emitting electrons are usually less energetic than 
X-ray-emitting ones, they do not suffer from severe radiative losses and the majority of them may survive 
from (and hence probe) early epochs of the PWN evolution. This interpretation has been further supported 
by the discovery of the spectral softening of the VHE nebula HESS~J1825-137 as a function of the 
distance from the pulsar \cite{c:j1825}. Given the discovery of this large population 
of middle-aged PWNe \cite{c:djannati09}, many new sources regularly revealed by the on-going \hess~Galactic 
Plane Survey could fall into this category, some of them being classified as PWN {\it candidates}. For 
instance, HESS~J1356-645 \cite{c:pwne08} lies close to the young ($\tau_{c}$ = 7.3 kyr) and energetic 
(\edot~= 3.1 $\times$ 10$^{36}$ erg s$^{-1}$) 166\un{ms} pulsar PSR~J1357-6429 \cite{c:camilo04}, for which 
only a marginal evidence of a 3\s~diffuse \xray~emission (\ie of a putative PWN) was 
found \cite{c:esposito07,c:zavlin07}. Interestingly, the extended VHE $\gamma$-ray source coincides with a diffuse 
radio emission, originally catalogued as a SNR candidate \cite{c:duncan97}. The on-going analysis of archival 
radio and \xray~data should thus help to constraint the nature of HESS~J1356-645 (Aharonian et al., in 
prep.). Therefore, these sources can be confirmed as VHE $\gamma$-ray PWNe thanks to a detailed investigation 
of all the available multi-wavelength data, together with follow-up observations in radio (\eg HESS~J1857+026
- PSR~J1856+0245) \cite{c:dark08,c:hessels08} and with now \fermi/LAT (\eg MGRO~J1908+06 - HESS~J1908+063 - 
0FGL~J1907.5+0617) \cite{c:milagro07,c:j1908,c:fermi09}, in order to reveal the associated (presumably 
energetic) pulsars.

\section{VHE $\gamma$-ray ``nebulae''}
\label{s:dark}

From the well-identified cases such as shell-type SNRs and PWNe, to the PWN candidates for which further data 
are required to firmly establish the putative association, it is presented here some cases of VHE $\gamma$-ray 
nebulae, also called dark sources \cite{c:dark08}. HESS~J1731-347 represents the best example in this regard. 
Originally classified as a dark source, a faint and extended (R $\sim$ 0.25\d) non-thermal radio and \xray~shell-type 
SNR, named G353.6-0.7, coincident with the extended VHE $\gamma$-ray emission, was discovered in the archival 
data \cite{c:tian08}. At a distance of $\sim$ 3.2\un{kpc}, estimated from HI absorption measurements 
toward an adjacent HII region, this SNR would have a physical diameter of $\sim$ 28\un{pc}, significantly 
larger than the known VHE shell-type SNRs described in section \ref{s:snrs}. This would then suggest that 
G353.6-0.7 is an old ($\sim$ 2.7 $\times$ 10$^{4}$ yrs) and intrinsically very bright SNR. However, 
its distance and the nature of the \xray~emission are poorly constrained. Even though some theoretical studies have 
proposed that old SNRs ($\sim$ 10$^{4-5}$ yrs) could still emit in the VHE $\gamma$-ray domain \cite{c:yamazaki06}, 
it is commonly thought that multi-TeV particles usually leave the acceleration site on timescales of a few 
thousands of years \cite{c:ptuskin05}. Therefore, follow-up radio and \xray~observations are needed in order 
to shed further light on the nature of this newly discovered shell-type SNR. 

This example serves as a discussion about the VHE $\gamma$-ray emission from SNRs. During the Sedov phase of the 
SNR evolution, accelerated particles are gradually injected in the ISM, the most energetic ones being released first. 
In case the SNR lies close to a molecular cloud (MC, at $\lesssim$ 100\un{pc}), delayed VHE $\gamma$-ray (and neutrino)
emission of the latter, through p-p interactions, may arise at detectable levels with the current IACTs 
\cite{c:gabici07}. The duration of VHE $\gamma$-ray emission from the cloud ($\gtrsim$ 10$^{4}$ yrs) would then last 
much longer than that of the SNR itself, since it is determined by the time of propagation of CRs from the accelerator 
to the target. On a theoretical side, the detection of such emission would then indicate that the nearby SNR was acting 
{\it in the past} as an effective Galactic CR accelerator or PeVatron \footnote{Thus, the cutoff measured in 
\rxj~at $\sim$ 20\un{TeV} \cite{c:j1713_3}, which translates into an E$_{max}$ of particles at $\gtrsim$ 100\un{TeV} 
(well below the knee in the CR spectrum observed at Earth), would imply that \rxj~{\it was a PeVatron in the past} 
and that the highest energy CRs have already been released in the surrounding medium}. On an observational side, in
this recently revisited scenario \cite{c:gabici07}, (some of) the unidentified VHE $\gamma$-ray sources could be 
{\it indirectly} associated with old SNRs, the $\gamma$-ray emission being produced during the interaction of escaping 
CRs with nearby MCs. One would then expect a correlation between the VHE $\gamma$-ray emission and the tracers of 
molecular material ($^{12}$CO, $^{13}$CO and masers in case the SNR shock encounters the MC), as it might be the case 
for the VHE $\gamma$-ray emission detected by \hess~toward the old SNRs W41 (HESS~J1834-087 \cite{c:gps06}) and W28 
(HESS~J1800-240/J1801-233 \cite{c:w28}), the CTB 37 complex \cite{c:ctb37A,c:ctb37B}, and HESS~J1745-303 
\cite{c:j1745}. However, it is worth noting that some of these nebulae could be instead VHE $\gamma$-ray PWNe 
\cite{c:mukherjee09}, as discussed in section \ref{s:plerion}: the large lifetime of VHE $\gamma$-ray emitting electrons 
in low magnetic field environments ($\sim$ 20 B$^{-2}_{\textrm{5}\mu\textrm{G}}$ E$^{-1/2}_{\gamma,\textrm{TeV}}$ kyr) 
makes the ratio of the VHE $\gamma$-ray luminosity to the \xray~luminosity an increasing function of the source age 
and size \cite{c:mattana09}. Therefore, one would expect VHE $\gamma$-ray PWNe to be hardly detectable with current 
\xray~instruments, and more generally at any other wavelength (leading to a VHE{\it -only}, dark, source). Moreover, 
for most of these nebulae, the morphology is poorly characterized, where only the source barycenter and gaussian 
extension are usually provided. Many of them may well be multiple sources of different kinds, as for HESS~J1800-240 
(A, B, \& C) \cite{c:w28} and HESS~J1745-303 \cite{c:j1745}. Observations with the next generation of IACTs such as CTA
and AGIS, with better sensitivities and angular resolutions, will undoubtedly help to search for counterparts with 
small field-of-view instruments and, thus, constrain the nature of the VHE $\gamma$-ray emission(s).

\section{VHE Galactic diffuse emission... or unresolved sources?}
\label{s:diffuse}

\begin{figure}[!htb]

\centering
\psfig{figure=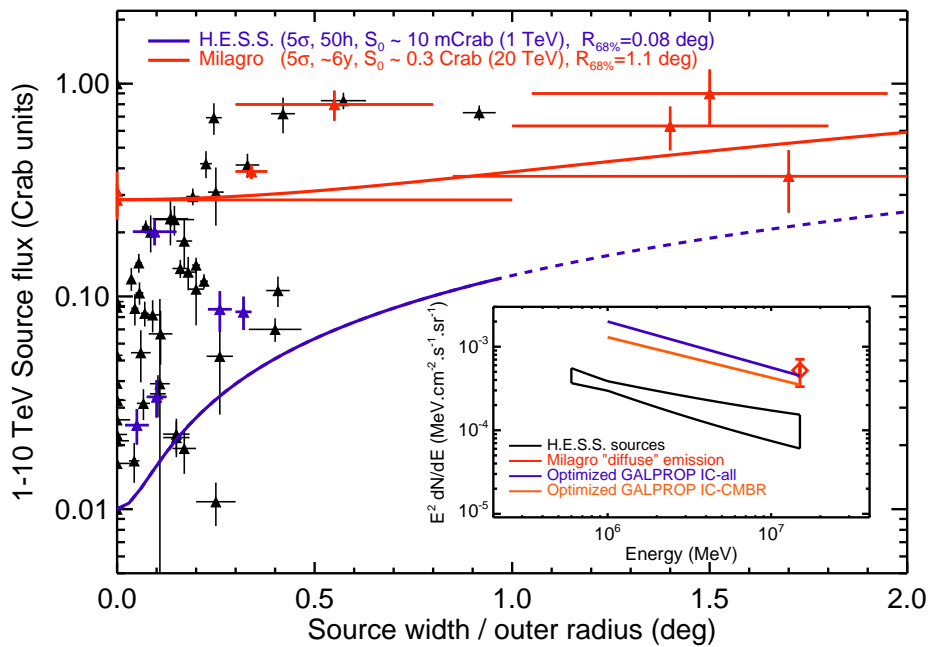,width=12.5cm}

\caption{Flux -- Size plot of the VHE Galactic sources detected so far. The integrated
flux is given between 1 and 10\un{TeV}, in units of the Crab nebula in the
same energy band. The size correponds to either the intrinsic mean source width or, in the
case of shell-type SNRs, the outer radius, in units of degrees. The red points represent 
the sources detected by \milagro~during its Galactic Plane survey ($\ell$ $\in$ [30,65]\d, 
$|$b$|$ $<$ 2\d), while the blue points correspond to those revealed by \hess~in the 
same region of the sky. The solid lines show the sensitivities of the two instruments, 
according to their respective characteristics given in brackets, which degrades with the
source extent as: S = S$_{0}$ $\times$ (R$_{s}^{2}$ + $\sigma_{psf}^{2}$)$^{1/2}$/$\sigma_{psf}$, 
where S$_{0}$ is the nominal sensitivity and R$_{s}$ is the effective source size. Note 
that this law does not hold anymore when the source size becomes comparable to the instrument 
field of view, which explains why the \hess~sensitivity curve is valid only for R$_{s}$ $\lesssim$ 
1\d. The inset plot shows the Milagro diffuse emission measured at 15\un{TeV}, the expected 
diffuse flux from the optimized GALPROP model and the summed spectrum of all the \hess~sources 
falling into the region probed by \milagro. \label{f:ima3}}

\end{figure}

This section is devoted to the recent detection by \milagro~of a large-scale VHE $\gamma$-ray diffuse emission 
(30\d~$<$ $\ell$ $<$ 110\d~and 136\d~$<$ $\ell$ $<$ 216\d, $|$b$|$ $<$ 10\d) \cite{c:milagro08}, after removing 
the contribution of the sources detected by this experiment \cite{c:milagro07}. In the inner part of the 
Galaxy (between 30 and 65\d~in longitude, \ie excluding the peculiar Cygnus region), the traditional 
GALPROP model fails to fit the measured flux at $\sim$ 15\un{TeV}. An optimized version of GALPROP has been 
designed to reproduce the \egret~data by relaxing the restriction of the local CR measurements. In this 
model, the electron spectrum is contrained by the \egret~data themselves, such that any hard and faint 
(relatively to the standard \pidec~spectrum from CR protons) IC spectrum, as proposed by the authors, would 
not violate the GeV measurements, while explaining the measured flux at 15\un{TeV}. Diffuse emission would 
thus be almost entirely explained by $\sim$ 100\un{TeV} electrons scattering off the CMB, with a flux, after 
propagation, of four times the one measured locally. Interestingly, this region has also been surveyed by 
\hess, featuring much better sensitivity and angular resolution than \milagro~at the expense of a much 
smaller field of view (see Figure \ref{f:ima3}), though with a non-uniform coverage \cite{c:gps08}. So far, 
five VHE $\gamma$-ray sources have been detected by \hess~(and unresolved by \milagro) in this region, and the 
resulting summed spectrum is shown in Figure \ref{f:ima3}, together with the \milagro~measurement. Roughly 20 \% 
of the diffuse emission is already explained by these \hess~sources, and a larger fraction should be 
reached in a next future once the existing \hess~data will be carefully analyzed and the survey will become 
uniform. First studies of the VHE $\gamma$-ray source population, based on the second \hess~survey catalogue 
\cite{c:gps06}, had already suggested that at least 10 \% of the VHE Galactic diffuse emission should be 
attributed to unresolved sources \cite{c:casanova08}. 

\begin{figure}[!htb]

\centering
\psfig{figure=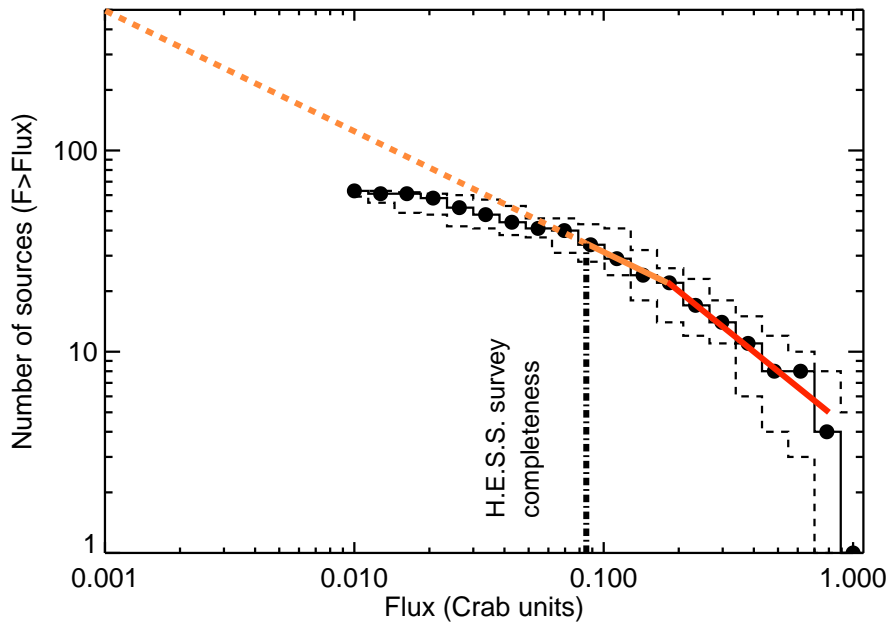,width=12cm}

\caption{Log N($>$S) -- log S diagram of the VHE Galactic sources. The dashed lines represent
the dispersion of the distribution after taking into account the statistical and systematical
errors on the source spectra through a Monte-Carlo simulation. The red and orange solid
lines correspond to the expectations from a uniform distribution in a thin disk (slope of -1)
and a source population distributed along the spiral arms (slope of -0.6), respectively. The
transition between these two regimes was set arbitrarly. The dot-dashed line shows the 
completeness limit of the \hess~survey (for a source extent of 0.2\d~in radius, and five hours 
of observations everywhere in the inner part of the Galaxy). The extrapolation down to 1 mCrab, 
a sensitivity that should be reached by the next generation of IACTs like CTA and AGIS, 
would then lead to the detection of about 500 VHE Galactic sources. \label{f:ima4}}

\end{figure}

\section{Conclusion}
\label{s:conclu}

To conclude this review, the log N($>$S) -- log S distribution of all the VHE Galactic sources known so far is 
shown in Figure \ref{f:ima4}. A slope of -1 corresponds to a uniform infinite plane distribution while a 
slope of $\sim$ -0.6 indicates that the population follows the spiral arm distribution on Galactic scales (\ie 
at distances of $\gtrsim$ 8\un{kpc}), as expected from young stellar population generating 
SNRs and PWNe. Even though there are inherent complications in the interpretation of such plot, the transition 
between the two regimes (at $\sim$ 0.15 Crab) seems to take place above the completeness limit of the \hess~survey. 
Extrapolating the curve down to 1 mCrab in sensitivity then implies that the next generation of IACTs like 
CTA \cite{c:cta} and AGIS \cite{c:agis} should detect at least 500 sources. As discussed in section \ref{s:dark}, 
many faint sources could actually be several sources, not clearly resolved yet, leading the log N($>$S) -- log S 
to soften toward low fluxes. The paper focused mainly on the latest results and open questions 
related to shell-type SNRs and PWNe. Besides them, new classes of VHE $\gamma$-ray emitters in the Milky-Way are 
expected to emerge, either from multi-wavelength follow-up observations of the so-called dark sources, as it 
might be the case of HESS~J1503-582 \cite{c:j1503}, or through dedicated observations of sky regions of interest, 
as exemplified by the recent detection of VHE $\gamma$-rays towards the young stellar cluster Westerlund~2 
\cite{c:wd2}. Such detection has triggered many exciting questions on the nature of the VHE $\gamma$-ray emission 
and more generally on the contribution of such wind-blown bubbles to the Galactic CR flux. HESS~J1848-018 could 
be the second case of this kind, where the VHE $\gamma$-ray emission is found to be slightly offset from the 
massive star-forming (Galactic ``mini-starburst'') region W~43 \cite{c:w43}. Moreover, $\gamma$-ray binaries 
have now joined the club of VHE $\gamma$-ray emitters and many pending issues still need to be investigated. Next 
generation of VHE observatories, in tight link with incoming multi-wavelength instruments, will undoubtedly 
play a crucial role in the understanding of all these acceleration sites in the Galaxy and, very likely, in the 
discovery of an even larger diversity of VHE $\gamma$-ray sources than expected.


\section*{References}
\begin{thebibliography}{99}

\small

\bibitem{c:duncan97}   A.R. Duncan \etal, \Journal{\mnras}{287}{722}{1997}
\bibitem{c:blondin01}  J.M. Blondin, R.A. Chevalier \& D.M. Frierson, \Journal{\apj}{563}{806}{2001}
\bibitem{c:slane01}    P. Slane \etal, \Journal{\apj}{548}{814}{2001}
\bibitem{c:casA01}     F.A. Aharonian \etal (\hegra~collaboration), \Journal{\aap}{370}{112}{2001}
\bibitem{c:tycho01}    F.A. Aharonian \etal (\hegra~collaboration), \Journal{\aap}{373}{292}{2001}
\bibitem{c:malkov01}   M.A. Malkov \& L.O.C. Drury, \Journal{RPPh}{64}{429}{2001}
\bibitem{c:camilo04}   F. Camilo \etal, \Journal{\apj}{611}{L25}{2004}
\bibitem{c:vds04}      E. van der Swaluw, T.P. Downes \& R. Keegan, \Journal{\aap}{420}{937}{2004}
\bibitem{c:bell04}     A.R. Bell, \Journal{\mnras}{353}{550}{2004}
\bibitem{c:j1713_1}    F.A. Aharonian \etal (\hess~collaboration), \Journal{\nat}{432}{75}{2004}
\bibitem{c:vbk05}      H.J. V\"olk, E.G. Berezhko \& L.T. Ksenofontov, \Journal{\aap}{433}{229}{2005}
\bibitem{c:pohl05}     M. Pohl, H. Yan \& A. Lazarian, \Journal{\apj}{626}{L101}{2005}
\bibitem{c:katagiri05} H. Katagiri \etal (\cangaroo~collaboration), \Journal{\apj}{619}{L163}{2005}
\bibitem{c:ptuskin05}  V.S. Ptuskin \& V.N. Zirakashvili, \Journal{\aap}{429}{755}{2005}
\bibitem{c:gps05}      F.A. Aharonian \etal (\hess~collaboration), \Journal{Science}{307}{1938}{2005}
\bibitem{c:g09}        F.A. Aharonian \etal (\hess~collaboration), \Journal{\aap}{432}{L25}{2005}
\bibitem{c:msh1552}    F.A. Aharonian \etal (\hess~collaboration), \Journal{\aap}{435}{L17}{2005}
\bibitem{c:sn1006_1}   F.A. Aharonian \etal (\hess~collaboration), \Journal{\aap}{437}{135}{2005}
\bibitem{c:velajr1}    F.A. Aharonian \etal (\hess~collaboration), \Journal{\aap}{437}{L7}{2005}
\bibitem{c:gs06}       B.M. Gaensler \& P.O. Slane, \Journal{\araa}{44}{17}{2006}
\bibitem{c:horns06}    D. Horns \etal, \Journal{\aap}{451}{L51}{2006}
\bibitem{c:parizot06}  E. Parizot \etal, \Journal{\aap}{453}{387}{2006}
\bibitem{c:porter06}   T.A. Porter, I.V. Moskalenko \& A.W. Strong, \Journal{\apj}{648}{L29}{2006}
\bibitem{c:enomoto06}  R. Enomoto \etal (\cangaroo~collaboration), \Journal{\apj}{652}{1268}{2006}
\bibitem{c:yamazaki06} R. Yamazaki \etal, \Journal{\mnras}{371}{1975}{2006}
\bibitem{c:gps06}      F.A. Aharonian \etal (\hess~collaboration), \Journal{\apj}{636}{777}{2006}
\bibitem{c:ridge}      F.A. Aharonian \etal (\hess~collaboration), \Journal{\nat}{439}{695}{2006}
\bibitem{c:velax}      F.A. Aharonian \etal (\hess~collaboration), \Journal{\aap}{448}{L43}{2006}
\bibitem{c:j1713_2}    F.A. Aharonian \etal (\hess~collaboration), \Journal{\aap}{449}{223}{2006}
\bibitem{c:crab}       F.A. Aharonian \etal (\hess~collaboration), \Journal{\aap}{457}{899}{2006}
\bibitem{c:j1825}      F.A. Aharonian \etal (\hess~collaboration), \Journal{\aap}{460}{365}{2006}
\bibitem{c:ellison07}  D.C. Ellison \etal, \Journal{\apj}{661}{879}{2007}
\bibitem{c:uchiyama07} Y. Uchiyama \etal, \Journal{\nat}{449}{576}{2007}
\bibitem{c:casA07}     J. Albert \etal, (\magic~collaboration) \Journal{\aap}{474}{937}{2007}
\bibitem{c:j1713_3}    F.A. Aharonian \etal (\hess~collaboration), \Journal{\aap}{464}{235}{2007}
\bibitem{c:velajr2}    F.A. Aharonian \etal (\hess~collaboration), \Journal{\apj}{661}{236}{2007}
\bibitem{c:wd2}        F.A. Aharonian \etal (\hess~collaboration), \Journal{\aap}{467}{1075}{2007}
\bibitem{c:twopwne}    F.A. Aharonian \etal (\hess~collaboration), \Journal{\aap}{472}{489}{2007}
\bibitem{c:milagro07}  A.A. Abdo \etal (\milagro~collaboration), \Journal{\apj}{664}{L91}{2007}
\bibitem{c:esposito07} P. Esposito \etal, \Journal{\aap}{467}{L45}{2007}
\bibitem{c:zavlin07}   V.E. Zavlin, \Journal{\apj}{665}{L143}{2007}
\bibitem{c:gabici07}   S. Gabici \& F.A. Aharonian, \Journal{\apj}{665}{L131}{2007}
\bibitem{c:acero07}    F. Acero, J. Ballet \& A. Decourchelle, \Journal{\aap}{475}{883}{2007}
\bibitem{c:dark08}     F.A. Aharonian \etal (\hess~collaboration), \Journal{\aap}{477}{353}{2008}
\bibitem{c:w28}        F.A. Aharonian \etal (\hess~collaboration), \Journal{\aap}{481}{401}{2008}
\bibitem{c:j1745}      F.A. Aharonian \etal (\hess~collaboration), \Journal{\aap}{483}{509}{2008}
\bibitem{c:j1913}      F.A. Aharonian \etal (\hess~collaboration), \Journal{\aap}{484}{435}{2008}
\bibitem{c:ctb37B}     F.A. Aharonian \etal (\hess~collaboration), \Journal{\aap}{486}{829}{2008}
\bibitem{c:kepler08}   F.A. Aharonian \etal (\hess~collaboration), \Journal{\aap}{488}{219}{2008}
\bibitem{c:ctb37A}     F.A. Aharonian \etal (\hess~collaboration), \Journal{\aap}{490}{685}{2008}
\bibitem{c:milagro08}  A.A. Abdo \etal (\milagro~collaboration), \Journal{\apj}{688}{1078}{2008}
\bibitem{c:enomoto08}  R. Enomoto \etal (\cangaroo~collaboration), \Journal{\apj}{683}{383}{2008}
\bibitem{c:tian08}     W.W. Tian \etal, \Journal{\apj}{679}{L85}{2008}
\bibitem{c:vkb08}      H.J. V\"olk, L.T. Ksenofontov \& E.G. Berezhko, \Journal{\aap}{490}{515}{2008}
\bibitem{c:bv08}       E.G. Berezhko \& H.J. V\"olk, \Journal{\aap}{492}{695}{2008}
\bibitem{c:butt08}     Y. Butt \etal, \Journal{\mnras}{386}{L20}{2008}
\bibitem{c:ellison08}  D.C. Ellison \& A. Vladimirov, \Journal{\apj}{673}{L47}{2008}
\bibitem{c:plaga08}    R. Plaga, \Journal{New Astronomy}{13}{73}{2008}
\bibitem{c:amato08}    E. Amato, P. Blasi \& S. Gabici, \Journal{\mnras}{385}{1946}{2008}
\bibitem{c:zp08}       V.N. Zirakashvili \& V.S. Ptuskin, \Journal{\apj}{678}{939}{2008} 
\bibitem{c:bykov08}    A.M. Bykov, Y.A. Uvarov \& D.C. Ellison, \Journal{\apj}{689}{L133}{2008} 
\bibitem{c:hessels08}  J.W.T. Hessels \etal, \Journal{\apj}{682}{L41}{2008}
\bibitem{c:casanova08} S. Casanova \& B.L. Dingus, \Journal{APh}{29}{63}{2008}
\bibitem{c:youngpwne}  A. Djannati-Ata\"i \etal (\hess~collaboration), \Journal{Proceedings of the 30th ICRC}{2}{823}{2008}
\bibitem{c:w43}        R.C.G. Chaves \etal (\hess~collaboration), \Journal{AIP Proceedings of the 4th $\gamma_{\textrm{2008}}$}{1085}{372}{2008}
\bibitem{c:gps08}      R.C.G. Chaves \etal (\hess~collaboration), \Journal{AIP Proceedings of the 4th $\gamma_{\textrm{2008}}$}{1085}{219}{2008}
\bibitem{c:pwne08}     M. Renaud \etal (\hess~collaboration), \Journal{AIP Proceedings of the 4th $\gamma_{\textrm{2008}}$}{1085}{285}{2008}
\bibitem{c:j1503}      M. Renaud \etal (\hess~collaboration), \Journal{AIP Proceedings of the 4th $\gamma_{\textrm{2008}}$}{1085}{281}{2008}
\bibitem{c:kes75}      R. Terrier \etal (\hess~collaboration), \Journal{AIP Proceedings of the 4th $\gamma_{\textrm{2008}}$}{1085}{316}{2008}
\bibitem{c:reynolds08} S.P. Reynolds, \Journal{\araa}{46}{89}{2008}
\bibitem{c:gallant08}  Y. Gallant \etal, \Journal{AIP Conference Proceedings of ``40 years of pulsars''}{983}{195}{2008}
\bibitem{c:dejager08}  O.C. de Jager \& A. Djannati-Ata\"i, \Journal{AIP Proceedings of ``Neutron Stars and Pulsars: 40 years after their discovery''~in press}{--}{--}{2008}
\bibitem{c:djannati09} A. Djannati-Ata\"i, \Journal{NIMA}{602}{28}{2009}
\bibitem{c:rcw86}      F.A. Aharonian \etal (\hess~collaboration), \Journal{\apj}{692}{1500}{2009}
\bibitem{c:drury09}    L.O.C Drury \etal, \Journal{\aap}{496}{1}{2009}
\bibitem{c:patnaude09} D.J. Patnaude, D.C. Ellison \& P. Slane, \Journal{\apj~in press}{--}{--}{2009}
\bibitem{c:morlino09}  G. Morlino, E. Amato \& P. Blasi, \Journal{\mnras}{392}{240}{2009}
\bibitem{c:mattana09}  F. Mattana \etal, \Journal{\apj}{694}{12}{2009}
\bibitem{c:mukherjee09}R. Mukherjee, E.V. Gotthelf \& J.P. Halpern, \Journal{\apj}{691}{1707}{2009} 
\bibitem{c:fermi09}    A.A. Abdo \etal (\fermi~collaboration), \Journal{Submitted to \apjs}{--}{--}{2009}
\bibitem{c:j1908}      F.A. Aharonian \etal (\hess~collaboration), \Journal{Accepted in \aap}{--}{--}{2009}
\bibitem{c:sn1006_2}   M. Naumann-Godo \etal (\hess~collaboration), \Journal{These proceedings}{--}{--}{2009}
\bibitem{c:agis}       W. Benbow \etal, \Journal{These proceedings}{--}{--}{2009}
\bibitem{c:cta}        M. Punch \etal, \Journal{These proceedings}{--}{--}{2009}

\end{thebibliography}
\end{document}